\documentclass[prb,twocolumn]{revtex4}
\usepackage{amssymb}
\usepackage{graphicx}
\usepackage{amsthm}

\def\c#1{\mathbb{#1}}

\def\ket#1{\left|#1\right\rangle}
\def\tx#1{{\rm{#1}}}

\begin{document}
\title{
Tensor-product representations for string-net condensed states}
\author{
Zheng-Cheng Gu$^\dagger$, Michael Levin$^{\dagger\dagger}$, Brian
Swingle$^{\dagger}$, and  Xiao-Gang Wen$^{\dagger}$ } \affiliation{
Department of Physics, Massachusetts Institute of
Technology, Cambridge, Massachusetts 02139, USA$^{\dagger}$ \\
Department of Physics, Harvard University, Cambridge, Massachusetts
02138, USA $^{\dagger\dagger}$ }

\begin{abstract}
We show that general string-net condensed states have a natural
representation in terms of tensor product states (TPS) .  These
TPS's are built from local tensors.  They can describe both states
with short-range entanglement (such as the symmetry breaking states)
and states with long-range entanglement (such as string-net
condensed states with topological/quantum order).  The tensor
product representation provides a kind of 'mean-field' description
for topologically ordered states and could be a powerful way to
study quantum phase transitions between such states. As an attempt
in this direction, we show that the constructed TPS's are
fixed-points under a certain wave-function renormalization group
transformation for quantum states.
\end{abstract}

\maketitle

\section{Introduction}

In modern condensed matter theory, an essential problem is the classification
of phases of matter and the associated phase transitions. Long range
correlation and broken symmetry\cite{L3726} provide the conceptual foundation
to the traditional theory of phases. The mathematical description of such a
theory is realized very naturally in terms of order parameters and group
theory.  The Landau symmetry breaking theory was so successful that people
started to believe that the symmetry breaking theory described all phases and
phase transitions. From this point of view, the discovery of the fractional
quantum Hall (FQH) effect\cite{TSG8259} in the 1980's appears even more
astonishing than what we realized.  These unique phases of matter have taught
us a very important lesson, when quantum effects dominate, entirely new kinds
of order, orders not associated with any symmetry, are possible.\cite{Wrig}
Similarly, new type of quantum phase transitions, such as the continuous phase
transitions between states with the same
symmetry\cite{WWtran,SMF9945,Wctpt,Wqoslpub} and incompatible
symmetries\cite{SBS0407,SenthilQCP} are possible.  There is literally a whole
new world of quantum phases and phase transitions waiting to be explored.  The
conventional approaches, such as symmetry breaking and order parameters,
simply does not apply here.

The particular kind of order present in the fractional quantum Hall
effect is known as topological order,\cite{Wrig} or more
specifically, chiral topological order because of broken parity and
time reversal (PT) symmetry. Topological phases which preserve
parity and time reversal symmetry are also
possible,\cite{RS9173,Wsrvb,SF0050,MS0181,SP0258,MSP0202,BFG0212,Wqoslpub,Wqoexct,K032,IFI0203}
and we will focus on these phases in this paper.  An appealing
physical picture has recently been proposed for this large class of
PT symmetric topological phases in which the relevant degrees of
freedom are string like objects called string-nets.\cite{FNS0428,LWstrnet}
Just as particle condensation provides
a physical picture for many symmetry breaking phases, the physics of
highly fluctuation strings, string-net condensation, has been found
to underlie PT symmetric topological phases.

The physical picture of string-net condensation provides also a natural
mathematical framework, tensor category theory, which can be used to write
down fixed point wave functions and calculate topological quantum
numbers.\cite{LWstrnet,LWuni} These topological quantum numbers include the
ground state degeneracy on a torus, the statistics and braiding properties of
quasi-particles, and topological entanglement entropy.\cite{KP0604,LWtopent}
All these physical properties are quite non-local, but they can be studied in
a unified and elegant manner using the nonlocal string-net basis.

Unfortunately, this stringy picture for the physics underlying the topological
phase seems poorly suited for describing phase transitions out of the
topological phase.  The large and non-local string-net basis is difficult to
deal with in the low energy continuum limit appropriate to a phase transition.
Trouble arises from our inability to do mean field theory, to capture the
stringiness of the state in an average local way. Indeed, the usual toolbox
built around having local order parameters is no longer available since there
is no broken symmetry.  We are thus naturally led to look for a local
description of topological phases which lack traditional order parameters.

Remarkably, there is already a promising candidate for such a local
description. A new local ansatz, tensor product states (also called projected
entangled pair states), has recently been proposed for a large class of
quantum states in dimensions greater than one \cite{FrankPEPS1}. In the tensor
product state (TPS) construction and its generalizations,\cite{V0705,AV0804}
the wave function is represented by a local network of tensors giving an
efficient description of the state in terms of a small number of variables.

The TPS construction naturally generalizes the matrix product states in one
dimension.\cite{KSZ9155,PVW0701}  The matrix product state formulation
underlies the tremendous success of the density matrix renormalization group
for one dimensional systems.\cite{W9263} The TPS construction is useful for us
because it allows one to locally represent the patterns of long-range quantum
entanglement\cite{W0275,KP0604,LWtopent} that lies at the heart of topological
order. In this paper, we show that the general string-net condensed states 
constructed in \Ref{LWstrnet} have natural TPS
representations.  Thus the long-range quantum entanglement in a general
(non-chiral) topologically ordered state can be captured by TPS.  The local tensors that
characterize the topological order can be viewed as the analogue of the local
order parameter describing symmetry breaking order.

In addition to our basic construction, we demonstrate a set of invariance
properties possessed by the TPS representation of string-net states.  These
invariance properties are characteristic of fixed point states in a new
renormalization group for quantum states called the tensor entanglement
renormalization group.\cite{GLWtergV}  This fixed point 
property of string-net states has already been anticipated\cite{LWstrnet}, 
and provides a concrete demonstration of string-net states as the infrared 
limit of PT symmetric topological phases.

This paper is organized as follows. In sections 
\ref{Z2section}-\ref{semsection} we construct TPS representations for 
the two simplest string-net condensed states. In section \ref{gensection} 
we present the general construction. In the last section, we use 
the TPS representation to show that the string-net condensed states are 
fixed points of the tensor entanglement renormalization group. Most of 
the mathematical details can be found in the appendix.

\begin{figure}
\begin{center}
\includegraphics[width=1.8in]
{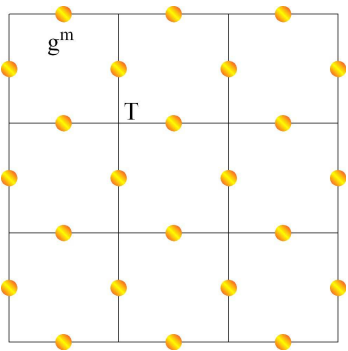}
\end{center}
\caption{$Z_2$ gauge model on a square lattice.  The dots represent
the physical states which are labeled by $m$.  The above graph can
also be viewed as a tensor-network, where each dot represents a
rank-3 tensor $g$ and each vertex represents a rank-4 tensor $T$.
The two legs of a dot represent the $\alpha$ and $\beta$
indices in the rank-3 tensor $g^m_{\alpha\beta}$.  The four legs
of a vertex represent the four internal indices in the rank-4
tensor $T_{\alpha\beta\gamma\lambda}$.
The indices on the connected links are summed over which define the
tensor trace tTr. } \label{Z2}
\end{figure}

\section{$Z_2$ gauge model}
\label{Z2section}

We first explain our construction in the case of the simplest string-net 
model: $Z_2$ lattice gauge theory (also known as the toric code). 
\cite{K032,Wqoexct,FNS0428,LWstrnet} In this
model, the physical degrees of freedom are spin-$1/2$ moments living on the
links of a square lattice.  The Hamiltonian is
\begin{eqnarray*}
 H= -\sum_p \prod_{i\in p} \sigma^x_i -\sum_v \prod_{i\in v} \sigma^z_i .
\end{eqnarray*}
Here $\prod_{i\in p}\sigma^x_i$ is the product of the four $\sigma^x_i$ around
a square $p$ and $\sum_p$ is a sum over all the squares.  The term $\prod_{i\in
v}\sigma^z_i$ is the product of the four $\sigma^z_i$ around a vertex $v$ and
$\sum_v$ is a sum over all the vertices.  The ground state $\ket{\Psi_{Z_2}}$ of
$H$ is known exactly.  To understand this state in the string language, we
interpret the $\sigma^z = -1$ and $\sigma^z = 1$ states on a single link as
the presence or absence of a string.  (This string is literally an electric
flux line in the gauge theory.) The ground state is simply an equal superposition of all closed
string states (e.g. states with an even number of strings incident at each vertex):
\begin{eqnarray}
\ket{\Psi_{Z_2}}=\sum_{X \rm{closed}} \ket{X},
\end{eqnarray}
While this state is relatively simple, it contains nontrivial topological order.
That is, it contains quasiparticle excitations with nontrivial statistics
(in this case, Fermi statistics and mutual semion statistics) and it
exhibits long range entanglement (as indicated by the non-zero topological entanglement 
entropy \cite{KP0604,LWtopent}). 

The above state has been studied before using TPS
\cite{FrankPEPS2,VidalMERA}.  Our TPS construction is different from
earlier studies because it is derived naturally from the string-net
picture. As illustrated in Fig. \ref{Z2}, we introduce two sets of
tensors: $T$-tensors living on the vertices and $g$-tensors living
on the links.  The $g$-tensors are rank three tensors
$g^m_{\alpha\alpha^\prime}$ with one physical index $m$ running over the two
possible spin states $\uparrow,\downarrow$, and two ``internal"
indices $\alpha, \alpha'$ running over some range $0,1,...,k$. The 
$T$-tensors are rank four tensors $T_{\alpha\beta\gamma\delta}$ with four 
internal indices $\alpha, \beta, \gamma, \delta$ running over
$0,1,...k$. 

In the TPS construction, we construct a quantum wave function
for the spin system from the two tensors $T$,$g$. The wave function is 
defined by
\begin{equation}
 \Psi(\{m_i\})=\text{tTr} [\otimes_v T \otimes_l g^{m_l}]
\end{equation}
To define the tensor-trace (tTr), one can introduce a graphic
representation of the tensors (see Fig. \ref{Z2}).  Then tTr means
summing over all unphysical indices on the connected links of tensor-network.

It is easy to check that the $Z_2$ string-net condensed ground state that 
we discussed above is given by the following choice of tensors with 
internal indices $\alpha, \beta...$ running over $0,1$:
\begin{eqnarray}
\label{T} T_{\alpha\beta\gamma\delta}=\left\{\begin{array}{cc}  1 &
{\rm{if}} \quad \alpha+\beta+\gamma+\delta \quad \rm{even} \\ 0&
{\rm{if}} \quad \alpha+\beta+\gamma+\delta \quad \rm{odd}
\end{array}\right.
\end{eqnarray}
\begin{eqnarray}
\label{g} g^{\uparrow}_{00}=1, \ \ \ \ g^{\downarrow}_{11}=1, \ \ \
\ \text{ others}=0,
\end{eqnarray}
The interpretation of these tensors is straightforward. The rank-3
tensor $g$ behaves like a projector which essentially sets the
internal index equal to the physical index so that $\alpha=1$
represents a string and $\alpha=0$ represents no string. The meaning
of the tensor $T_{\alpha\beta\gamma\delta}$ is also clear: it just
enforces the closed string constraint, only allowing an even number
of strings to meet at a vertex.  

In this example, we have shown how to represent the simplest string-net 
condensed state using the TPS construction. We now explain how to extend 
this construction to the general case.

\section{Double-semion model}
\label{semsection}

Let us start by turning to a slightly less straightforward model
which illustrates some details necessary for our TPS construction
for general string-net states.  The model, which we call the double-semion
model, is a spin-1/2 model where the spins are located on the links of
the honeycomb lattice. The Hamiltonian is defined in Eq. (40) in Ref.
\onlinecite{LWstrnet}. Here we will focus on the ground state which
is known exactly. As in the previous example, the ground state can
be described in the string language by interpreting
the $\sigma^z = -1$ and $\sigma^z = 1$ states on a single link as
the presence or absence of a string. The ground state wave function
is a superposition of closed string states weighted by different
phase factors:
\begin{eqnarray}
\label{semWF} \ket{\Psi_{\text{dsemion}}}=\sum_{X \rm{closed}}
(-)^{n(X)}\ket{X}
\end{eqnarray}
where $n$ is the number of closed loops in the closed-string state $X$.
\cite{FNS0428,LWstrnet} As in the previous example, this state contains non-trivial
topological order. In this case, the state contains quasiparticle excitations
with semion statistics.
 
\begin{figure}
\begin{center}
\includegraphics[width=2.8in] {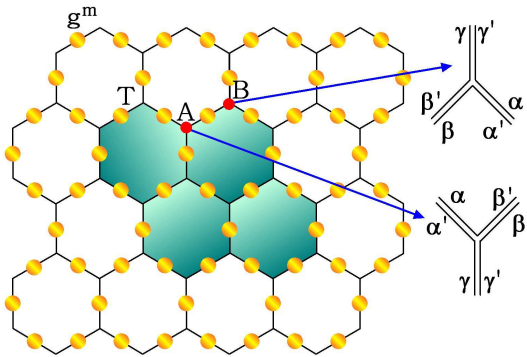}
\end{center}
\caption{The double-semion model on the honeycomb lattice. The
ground state wavefunction (\ref{semWF}) has a TPS representation
given by the above tensor-network.  Note that $T$ and $g$ has a
double line structure. Note that the vertices form a honeycomb
lattice which can divided into A-sublattice and B-sublattice. }
\label{semion}
\end{figure}

Like the $Z_2$ state, the above string-net condensed state can 
be written as a TPS with one set of tensors $T$ 
on the vertices and another set of tensors $g$ on links.
However, in this case it is more natural to use a
rank-6 tensor $T$ and a rank-5 tensor $g$. These tensors can be
represented by double lines as in Fig. \ref{semion}. The $T$-tensors
are given by
\begin{eqnarray}
\text{sublattice A}:\quad
T_{\alpha\alpha^\prime;\beta\beta^\prime;\gamma\gamma^\prime}&=&
T^0_{\alpha\beta\gamma}\delta_{\alpha\beta^\prime}
\delta_{\beta\gamma^\prime}\delta_{\gamma\alpha^\prime}\nonumber\\
\text{sublattice B}:\quad
T_{\alpha\alpha^\prime;\beta\beta^\prime;\gamma\gamma^\prime}&=&
T^0_{\alpha\beta\gamma}\delta_{\alpha^\prime\beta}
\delta_{\beta^\prime\gamma}\delta_{\gamma^\prime\alpha}\label{semion1}
\end{eqnarray}
where now each internal unphysical index $\alpha,\beta,...$ runs over $0,1$.
Here, the tensor $T^0$ is given by \begin{eqnarray}
T^0_{\alpha\beta\gamma} =\left\{\begin{array}{cc}
1 & {\rm{if}}\quad \alpha+\beta+\gamma=0,3\\
i & {\rm{if}}\quad \alpha+\beta+\gamma=1 \\
-i& {\rm{if}}\quad \alpha+\beta+\gamma=2
\end{array}\right.,
\label{semion2}
\end{eqnarray}
The rank-5 $g$ tensors are given by
\begin{eqnarray}
g^{\uparrow}_{00,00}=g^{\uparrow}_{11,11}=1, \ \
g^{\downarrow}_{01,01}=g^{\downarrow}_{10,10}=1, \ \ \text{
others}=0 . \label{semion3}
\end{eqnarray}
Again, the ground state wave function can be obtained by summing
over all the internal indices on the connected links in the tensor
network (see Fig. \ref{semion}):
\begin{eqnarray}
\label{dsemion} | \Psi_\tx{dsemion}\rangle
=\sum_{\{m_i\}}\tx{tTr}[\otimes_v T \otimes_l g^{m_i}]
|m_1,m_2,...\rangle  .
\end{eqnarray}

From Eq.(\ref{semion3}) we see that the physical indices and the
internal indices have a simple relation: each pair of internal unphysical
indices describes the presence/absence of string on the corresponding
link. Two identical indices (00 and 11) in a pair correspond to
no string (spin up) on the link and two opposite indices (01 and 10)
in a pair correspond to a string (spin down) on the link.  We may
think of each half of the double line as belonging to an associated
hexagon, and because every line along the edge of a hexagon takes
the same value, we can assign that value to the hexagon. In this way
we may view physical strings as domain walls in some fictitious
Ising model as indicated by the coloring in Fig. \ref{semion}.  The
peculiar assignment of phases in $T^0$ serve to guarantee the right
sign oscillations essentially by counting the number of left and
right turns made by the domain wall.

Equation (\ref{dsemion}) is interesting since the wave function
(\ref{semWF}) appears to be intrinsically non-local. We cannot
determine the number of closed loops by examining a part of a
string-net.  We have to examine how strings are connected in the
whole graph. But such a ``non-local'' wave function can indeed be
expressed as a TPS in terms of local tensors.

\section{General string-net models}
\label{gensection}

We now show that the general string-net condensed states
constructed in \Ref{LWstrnet} can be written naturally as TPS. 
To this end, we quickly review the basic properties of the
general string-net models and string-net condensed states. 

The general string-net models are spin models where the spins live 
on the links of the honeycomb lattice. Each spin can be in 
$N+1$ states labeled by $a=0,1,...,N$. The Hamiltonians
for these models are exactly soluble and are defined in  
Eq. (11) in Ref. \onlinecite{LWstrnet}. Here, we focus on
the (string-net condensed) ground states of these Hamiltonians.

In discussing these ground states it will be convenient
to use the string picture. In this picture, we regard
a link with a spin in state $a \neq 0$ as being occupied by a 
type-$a$ string. We think of a link with $a =0$ as being
empty. As in the previous examples, the ground states are 
superpositions of many different string configurations. However, in 
the more general case, the strings can branch (e.g. three strings can meet at
a vertex).

To specify a particular string-net model or equivalently
a particular string-net condensed state, one needs to
provide certain data. First, one needs to specify an integer 
$N$ - the number of string types. Second, one needs to give
a rank-3 tensor $\del_{abc}$ taking values $0,1$, where 
the indices $a,b,c$ range over $0,1,...,N$. This tensor
describes the branching rules: when $\del_{abc} = 0$,
that means that the ground state wave function does not
include configurations in which strings $a$, $b$, $c$ meet
at a point. On the other hand, if $\del_{abc} = 1$ then such
branchings are allowed. Third, the strings can have
an orientation and one needs to specify the string type $a^*$
corresponding to a type-$a$ string with the opposite orientation.
A string is not oriented if $a = a^*$. Finally, one needs
to specify a complex rank-6 tensor $F^{ijm}_{kln}$ where
$i,j,k,l,m,n=0,1,...,N$. The tensor $F^{ijm}_{kln}$ defines 
a set of local rules which implicitly define the wave 
function for the string-net condensed state. We would like to mention
that the data $(N,\del_{ijk}, F^{ijm}_{kln})$ cannot be specified 
arbitrarily. They must satisfy
special algebraic relations in order to define a valid
string-net condensed state.

The main fact that we will use in our construction of the TPS
representation of general string-net condensed states is that the 
string-net condensed states can be constructed by applying local projectors 
$B_p$ ($p$ is a plaquette of the honeycomb lattice) to a no-string state
$\ket{0}$.\cite{LWstrnet} These projectors $B_p$ can be written as
\begin{equation}
B_p = \sum_s a_s B^s_p 
\end{equation}
where $B^s_p$ has the simple physical meaning of adding a loop of type-$s$ 
string around the hexagon $p$. The constants $a_s$ are given by
\begin{equation}
a_s = \frac{d_s}{D}
\end{equation}
where $d_s = 1/F^{ss^*0}_{ss^*0}$, and $D = \sum_s d_s^2$.

This fact enables us to write the string-net condensed state as
\begin{eqnarray}
\label{strnet} \ket{\Psi_{\text{strnet}}}&=&\prod_pB_p\ket{0}
=\prod_p \sum_s a_s B_p^s \ket{0}
\nonumber\\
&=&\sum_{u,s,t,\cdots}a_ta_sa_u\cdots \ket{t,s,u,\cdots}
\end{eqnarray}
where
\begin{equation}
\ket{t,s,u,\cdots}_\text{coh} = B^t_{p_1} B^s_{p_2} B^u_{p_3}\cdots
\ket{0} .
\end{equation}
Note that $[ B^t_{p_1}, B^s_{p_2}]=0$ when $p_1\neq p_2$, and the
order of the $B^s_p$ operators in the above is not important. 

\begin{figure}
\begin{center}
\includegraphics[width=3in] {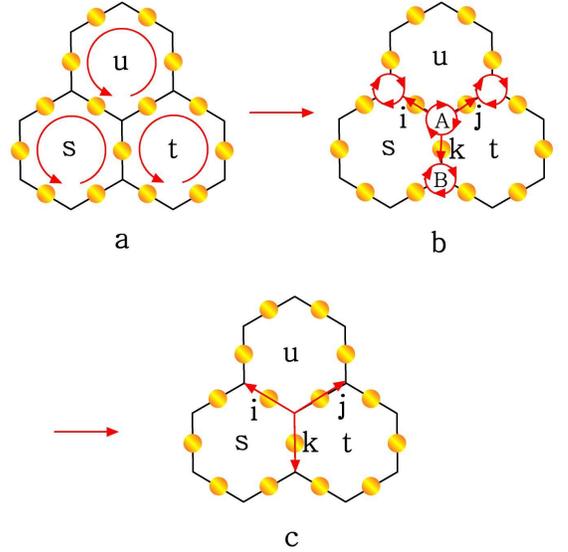}
\end{center}
\caption{ Using the fusion rules, we can represent the coherent
states $\ket{t,s,u,\cdots}$ in terms of the orthogonal string-net
states.} \label{stringnet}
\end{figure}

These states are not orthogonal to each other.  In the following, we would
like to express these states in terms of the orthonormal basis
of different string-net configurations:
\begin{eqnarray}
\ket{t,s,u,\cdots}_\text{coh} &=& \sum_{i,j,k,\cdots}
\Phi^{t,s,u,\cdots}_{i,j,k,\cdots} \ket{i,j,k,\cdots}
\end{eqnarray}
Here $\ket{i,j,k,\cdots}$ is a string-net configuration and $i,j,k,\cdots
=0,\cdots,N$ label the string types (\emph{i.e.} the physical
states) on the corresponding link (see Fig.  \ref{stringnet}c). We
note that $t,s,u,\cdots =0,\cdots,N$ label the string types
associated with the hexagons (see Fig. \ref{stringnet}a). As a
result, the string-net condensed state is given by
\begin{eqnarray*}
 \ket{\Psi_{\text{strnet}}}&=&
\sum_{t,s,u,\cdots} a_ta_sa_u\cdots \sum_{i,j,k,\cdots}
\Phi^{t,s,u,\cdots}_{i,j,k,\cdots} \ket{i,j,k,\cdots}
\end{eqnarray*}

\begin{figure}
\begin{center}
\includegraphics[scale=0.15] {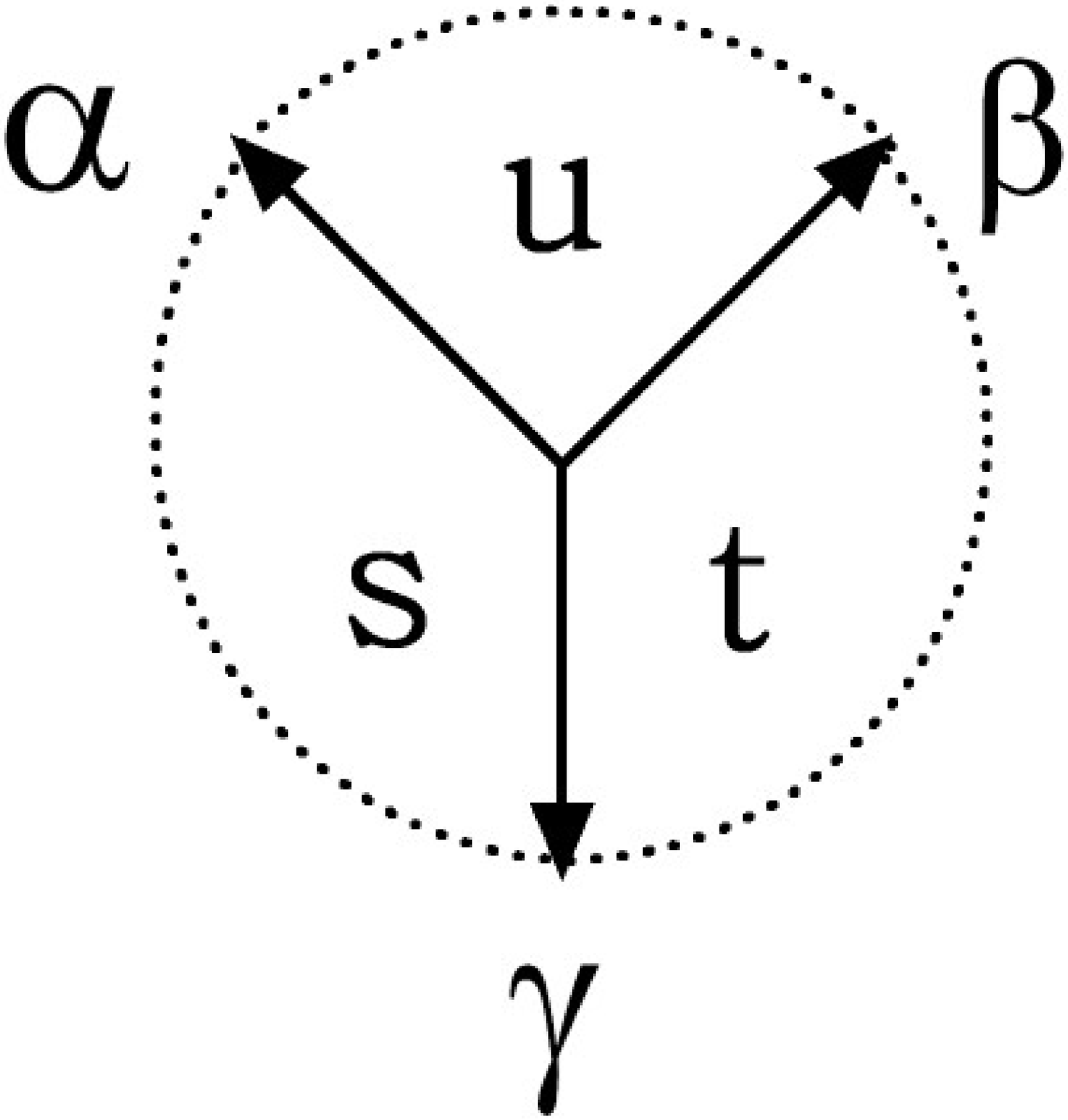}
\end{center}
\caption{ The graphic representation for the tensor
$G^{\al\bt\ga}_{tsu}$.} \label{graphicT0}
\end{figure}

To calculate $\Phi^{t,s,u,\cdots}_{i,j,k,\cdots}$, we note that the
coherent states $\ket{t,s,u,\cdots}_\text{coh}$ can be viewed a
string-net state in a fattened lattice (see Fig.
\ref{stringnet}a).\cite{LWstrnet}.  To obtain the string-net states
where strings live on the links, we need to combine the two strings
looping around two adjacent hexagons into a single string on the
link shared by the two hexagons.  This can be achieved by using the
string-net recoupling rules\cite{LWstrnet} (see Fig.
\ref{stringnet}).  This allows us to show that the string-net
condensed state can be written as (see Appendix \ref{rectps})
\begin{align}
\label{stringnetwave1} \ket{\Psi_{\text{strnet}}} &=&
\sum_{t,s,u,\cdots} \Big[ \prod_{\text{hexagon}} a_t  \Big]
\sum_{i,j,k,\cdots} \times
\\
&&\Big[ \prod_{\text{vert.}} \sqrt{v_iv_jv_k}G^{ijk}_{tsu}
\Big]\ket{i,j,k,..},
\nonumber
\end{align}
where $G^{ijk}_{tsu}\equiv F^{ijk}_{ts^*u}/(v_k v_u)$ is the
symmetric $6j$ symbol with full tetrahedral symmetry,\footnote{Such
a $G^{ijk}_{tsu}$ is slightly different from that introduced in
Ref. \onlinecite{LWstrnet}.} and $v_i=\sqrt{d_i}$.  

Let us explain the
above expression in more detail.  The indices $i,j,k,...$ are on the
links while the indices $u,t,s,...$ are on the hexagons.  Each
hexagon contributes to a factor $a_s$. Each vertex in A-sublattice
contributes to a factor $\sqrt{v_iv_jv_k}G^{ijk}_{tsu}$ and each
vertex in B-sublattice contributes to a factor
$\sqrt{v_iv_jv_k}G^{i^*j^*k^*}_{tsu}$.  The indices $i,j,k,t,s,u$
around a vertex are arranged as illustrated in Fig. \ref{graphicT0}.

\begin{figure}
\begin{center} \includegraphics[width=3.4in] {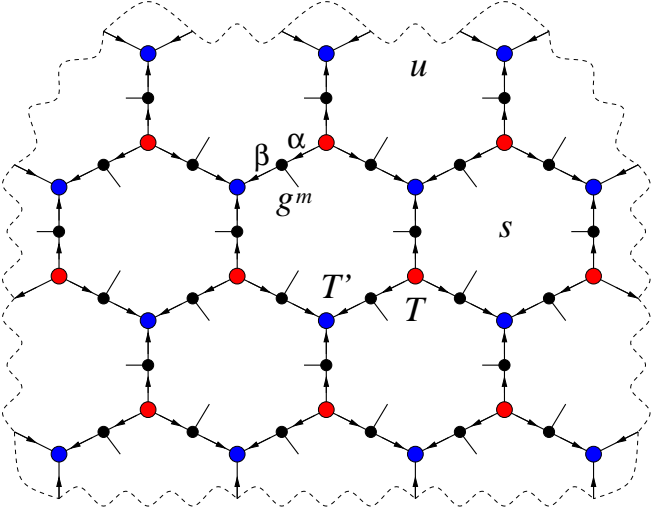}
\end{center}
\caption{ A tensor-complex formed by vertices, links, and faces. The
dashed curves are boundaries of the faces.  The links that connect
the dots carry index $\al,\bt,...$ and the faces carry index
$u,s,...$.  Each trivalent vertex represents a $T$-tensor.  The
vertices on A-sublattice (red dots) represents
$T_{\al\bt\ga;t,s,u}$. The vertices on B-sublattice (blue dots)
represents $T'_{\al\bt\ga;t,s,u}= T_{\al^*\bt^*\ga^*;t,s,u}$. The
dots on the links represent the $g^m$-tensor $g^m_{\al,\bt}$.  In
the weighted tensor trace, the $\al,\bt,...$ indices on the links
that connect the dots are summed over, and the $u,s,...$ indices on
the closed faces are summed over with a weighting factor $a_ua_s...$.
}
\label{hexgT}
\end{figure}

\begin{figure}
\begin{center}
\includegraphics[scale=0.6] {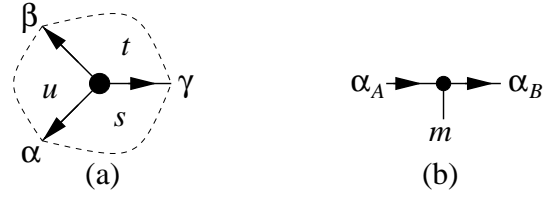}
\end{center}
\caption{ The graphic representation of (a) the $T$-tensor,
$T_{\al\bt\ga; tsu} $, and (b) the $g^m$-tensor $ g^m_{\al_A\al_B}
$.
} \label{gT}
\end{figure}

\begin{figure}
\begin{center}
\includegraphics[scale=0.55] {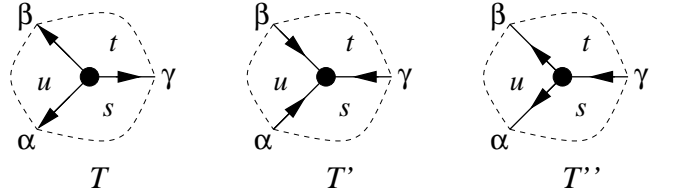}
\end{center}
\caption{ Three $T$-tensors associated with different orientations
of the legs are related: $T_{\al\bt\ga; tsu} $, $T'_{\al\bt\ga; tsu} = T_{\al^*\bt^*\ga^*; tsu} $, $T''_{\al\bt\ga; tsu} =
T_{\al\bt\ga^*; tsu} $.
} \label{TTT}
\end{figure}

The expression \eq{stringnetwave1} can be formally written as a
weighted tensor trace over a tensor-complex formed by $T$-tensors
and $g^m$-tensors. First, let us explain what is a tensor-complex. A
tensor-complex is formed by vertices, links, and faces (see Fig.
\ref{hexgT}).  The $T$-tensors live on the vertices and the $g^m$-tensor
live on the links.  The $T$-tensor carries the indices from
the three connected links and the three adjacent faces, while the
$g^m$ tensor carries the indices from the two connected links (see
Fig. \ref{gT}).  By having indices on faces, the tensor-complex
generalize the tensor-network.

The weighted tensor trace sums over all the $\al,\bt,...$ indices on the
internal links that connect two dots and sums over all the $u,s,...$ indices
on the internal faces that are enclosed by the links, with weighting factors
$a_u$, $a_s$, \etc from each enclosed faces.  Let us choose the $T$-tensors on
vertices to be (see Fig. \ref{TTT})
\begin{eqnarray}
\label{strcmpT}
\text{A-sublattice: }&& T_{\alpha\beta\gamma;tsu} =
\sqrt{v_\alpha v_\beta v_\gamma} G^{\alpha\beta\gamma}_{tsu},
\nonumber\\
\text{B-sublattice: }&& T'_{\alpha\beta\gamma;tsu} =
\sqrt{v_\alpha v_\beta v_\gamma} G^{\alpha^*\beta^*\gamma^*}_{tsu}.
\end{eqnarray}
and the $g^m$ tensor to be
\begin{eqnarray}
\label{strcmpg}
g^m_{\alpha_A \alpha_B }
&=&
\delta_{\alpha_A m}\delta_{\al_A\alpha_B}.
\end{eqnarray}
In this case the string-net condensed state \eq{stringnetwave1} can be
written as a weighted tensor trace over a tensor-complex:
\begin{eqnarray}
\label{strcmpWF1}
\ket{\Psi_{\text{strnet}}}=\sum_{m_1,m_2,...}\tx{wtTr}[\otimes_v T
\otimes_l g^{m_l}] |m_1,m_2,...\rangle  ,
\end{eqnarray}
where $m_i$ label the physical states on the links.  We note that
the $g^m$-tensor is just a projector: it makes each edge of the
hexagon to have the same index that is equal to $m$. The string-net
condensed state \eq{stringnetwave1} can also be written as a more
standard tensor trace over a tensor-network (see Appendix
\ref{strtnsnet}).

\begin{figure}
\begin{center}
\includegraphics[scale=0.6] {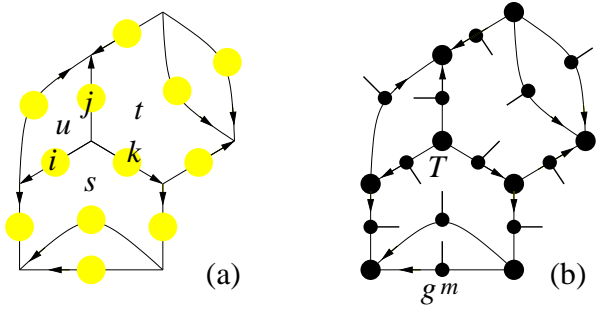}
\end{center}
\caption{ (a) The string-net condensed states on arbitrary trivalent
graph in two dimension can be written as (b) a weighted tensor trace
over a tensor-complex formed by $T$- and
$g^m$-tensor.
} \label{grph}
\end{figure}

The string-net states and the corresponding TPS representation can also be
generalized to \emph{arbitrary} trivalent graph in two dimension.  After a
similar calculation as that on the honeycomb lattice, we find that an
expression similar to Eq. (\ref{stringnetwave1}) describes the string-net
condensed state on a generic trivalent graph in two dimension (see Fig.
\ref{grph}a).  The indices in Eq. (\ref{stringnetwave1}), such as
$i,j,k,u,s,t,\cdots$, should be read from the Fig.  \ref{grph}a, where the
indices of the $G$-symbol is determined by three oriented legs and three faces
between them (see Fig. \ref{graphicT0}).

Such a string-net wavefunction
can be expressed in terms of weighted tensor-trace over a tensor-complex (see
Fig. \ref{grph}b):
\begin{eqnarray}
\label{strnetwtTr}
\ket{\Psi_{\text{strnet}}}=\sum_{m_1,m_2,...}\tx{wtTr}[\otimes_v T
\otimes_l g^{m_l}] |m_1,m_2,...\rangle  . \label{strnetWF1}
\end{eqnarray}
where the $T$-tensors, depending on the orientations on the legs, are given by
Fig. \ref{TTT} and  \eq{strcmpT}.

\section{The string-net wave function as a fixed point wave function}

An interesting property of the string-net condensed states constructed in
\Ref{LWstrnet} is that they have a vanishing correlation length: $\xi = 
0$. This suggests that the string-net condensed states are 
fixed points of some kind of renormalization group transformation. 
\cite{LWstrnet} In this section, we attempt to make this 
proposal concrete. We show that the string-net condensed states are 
fixed points of the tensor entanglement renormalization group (TERG) 
introduced in \Ref{GLWtergV}.

To understand and motivate the TERG, it is useful to first think about the 
problem of computing the norm and local density matrix of the string-net 
state (\ref{strnetWF1}). To obtain the norm, we note that
\begin{eqnarray}
&&\langle{u_2,s_2,t_2,\cdots}_\text{coh}|{u_1,s_1,t_1,\cdots}_\text{coh}\rangle\\
&=&\sum_{i,j,k,\cdots}\sum_{i^\prime,j^\prime,k^\prime,\cdots}
\cdots
F^{u_1^*u_10}_{s_1^*s_1i}F^{t_2^*t_20}_{u_1^*u_1j}F^{s_1^*s_10}_{t_1^*t_1k}\nonumber\\&&\times
\left(F^{u_2^*u_20}_{s_2^*s_2i^\prime}F^{t_2^*t_20}_{u_2^*u_2j^\prime}F^{s_2^*s_20}_{t_2^*t_2k^\prime}\right)^*
\cdots \langle X_b^\prime|X_b\rangle
\nonumber\\&=&\sum_{i,j,k,\cdots}\Big[ \prod_{\text{vertices}}
\frac{v_iv_jv_k}{v_{t_1}v_{s_1}v_{u_1}v_{t_2}v_{s_2}v_{u_2}}
F^{ijk}_{t_1s_1^*u_1} \left(F^{ijk}_{t_2 s_2^*
u_2}\right)^*\Big]\nonumber\\&&\times\frac{v_{s_1}v_{t_1}}{v_k}\frac{v_{s_2}v_{t_2}}{v_k}\nonumber\\&=&\sum_{i,j,k,\cdots}\Big[
\prod_{\text{vertices}} v_iv_jv_k G^{ijk}_{t_1s_1u_1}
G^{k^*j^*i^*}_{u_2 s_2 t_2}\Big]\nonumber
\end{eqnarray}
Thus, the norm can be written as
\begin{eqnarray}
\label{norm} && \langle \Psi_{\text{strnet}}|\Psi_{\text{strnet}}
\rangle = \sum_{i,j,k,\cdots}
\sum_{t_1,s_1,u_1,\cdots}\sum_{t_2,s_2,u_2,\cdots}\\ && \Big[
\prod_{\text{hexagon}} a_{t_1}a_{t_2}  \Big] \Big[
\prod_{\text{vertices}} v_iv_jv_k G^{ijk}_{t_1s_1u_1}
G^{k^*j^*i^*}_{u_2 s_2 t_2} \Big]
\nonumber\\
&=& \text{wtTr}\left[\c T\otimes \c T \otimes \c T
\otimes \c T \cdots \right] ,
\nonumber
\end{eqnarray}
where we have used the identity
${\left(F^{ijk}_{tsu}\right)}^*=F^{k^*j^*i^*}_{ust}\frac{v_kv_u}{v_iv_t}$.

\begin{figure}
\begin{center} \includegraphics[width=3.0in] {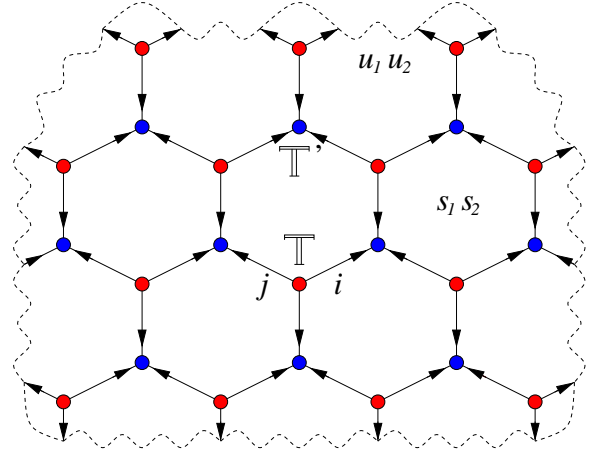}
\end{center}
\caption{ A tensor-complex with vertices, links, and faces is formed
by the double-tensors $\c T$ and $\c T'$.  (The dashed curves are
boundaries of the faces.) The links that connect the dots carry
index $i,j,...$ and the faces carry double-index
$u_1u_2,s_1s_2,...$.  Each trivalent vertex represents a double-$\c
T$-tensor.  The vertices on A-sublattice (red dots) represents $\c
T_{ijk;t_1t_2,s_1s_2,u_1u_2}$.  The vertices on B-sublattice (blue
dots) represents $\c T'_{ijk;t_1t_2,s_1s_2,u_1u_2}= \c
T_{i^*j^*k^*;t_1t_2,s_1s_2,u_1u_2}$.  In the weighted tensor trace,
the $i,j,...$ indices on the links that connect the dots are summed
over and the $u_1u_2,s_1s_2,...$ indices on the closed faces are
summed over independently with a weighting factor
$a_{u_1}a_{u_2}a_{s_1}a_{s2}...$.
} \label{hexDT}
\end{figure}

\begin{figure}
\begin{center}
\includegraphics[scale=0.35] {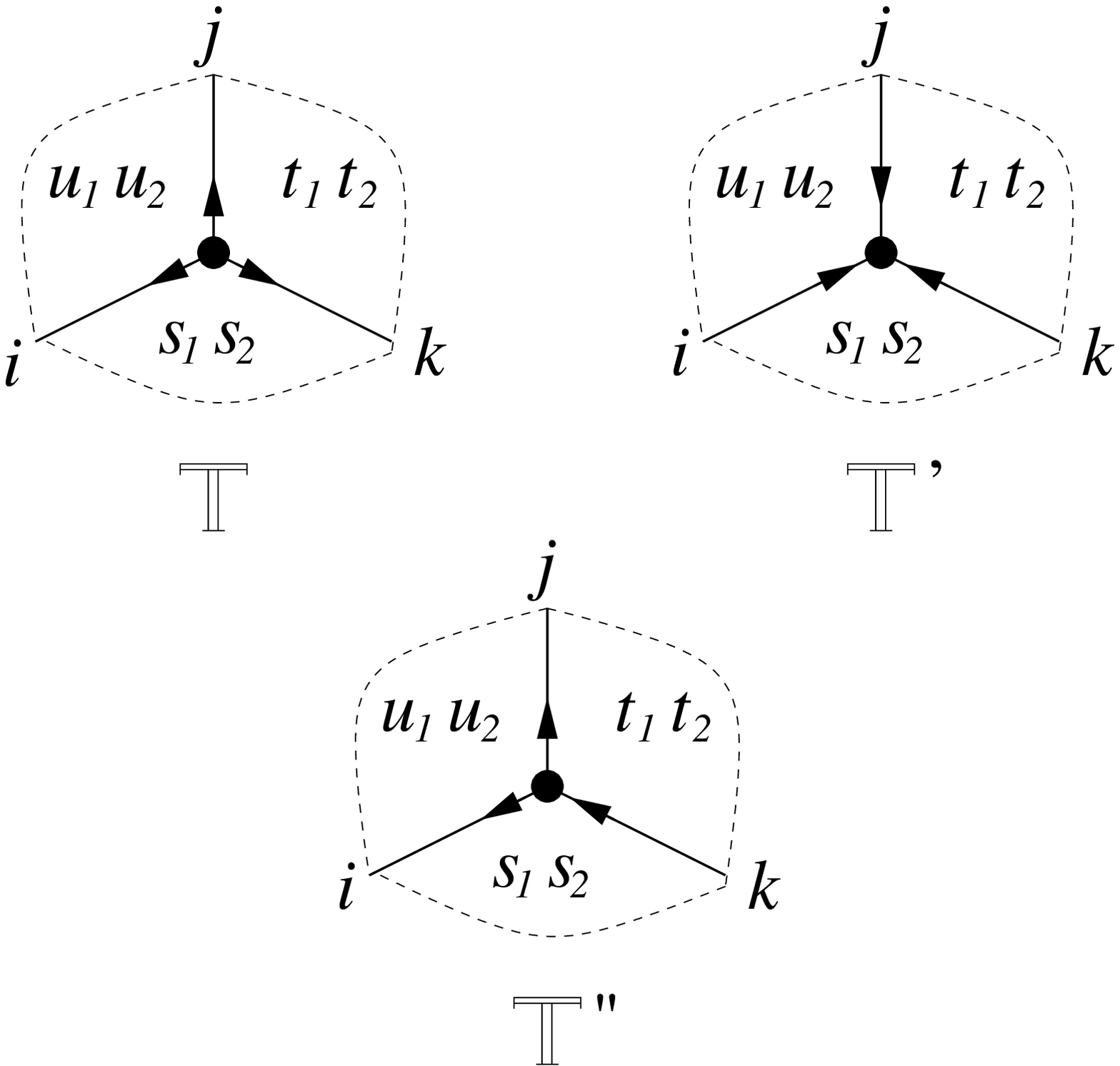}
\end{center}
\caption{ The graphic representations for the three double-$\c
T$-tensors $\c T_{ijk;t_1t_2,s_1s_2,u_1u_2}$, $\c
T^\prime_{ijk;t_1t_2,s_1s_2,u_1u_2} =\c
T_{i^*j^*k^*;t_1t_2,s_1s_2,u_1u_2}$, and $\c
T^{\prime\prime}_{ijk;t_1t_2,s_1s_2,u_1u_2} =\c
T_{ijk^*;t_1t_2,s_1s_2,u_1u_2}$.
} \label{DT}
\end{figure}

Here wtTr is a tensor trace over a tensor-complex formed by the double-tensor
$\c T $ (see Fig. \ref{hexDT}).  Again, the tensor-complex are formed by
vertices, links and faces.  Each trivalent vertex on the A(B)-sublattice
represent a double-tensor $\c T$ ($\c T'$):
\begin{align}
\label{DTeq} \text{A: } \c T_{ijk;s_1s_2,t_1t_2,u_1u_2} &=& v_i v_j
v_k G^{ijk}_{t_1s_1u_1}G^{k^*j^*i^*}_{u_2s_2t_2},
\nonumber\\
\text{B: }  \c T'_{ijk;s_1s_2,t_1t_2,u_1u_2} &=& v_i v_j v_k
G^{i^*j^*k^*}_{t_1s_1u_1}G^{kji}_{u_2s_2t_2},
\end{align}
The double-tensor $\c T$ can be represented by a graph with three
oriented legs and three faces between them (see in Fig. \ref{DT})
where each leg or face now carries two indices.  In wtTr, we sum
over all indices on the internal links.  We also sum over all
indices on the internal faces independently with weighting factors
$a_{u_1}a_{u_2}$ from each of the internal faces.  

With minor modification, this expression for the norm can be used to 
compute expectation values for local operators. Indeed, the local 
density matrix is given by a similar tensor trace, except that the
physical indices $i,j,k,...$ need to be left unsummed in the region where 
we want to compute the density matrix. Thus, the problem of computing 
expectation vales and norms can be reduced to the problem of evaluating 
the tensor trace (\ref{norm}). 

Unfortunately, evaluating tensor traces is an exponentially hard problem 
in two or higher dimensions. This leads us to the tensor entanglement 
renormalization group (TERG) method. 
The idea of the TERG method is to (approximately) evaluate tensor traces 
by coarse graining. \cite{LevinTRG} In each coarse graining step, the 
tensor complex $G$ with tensor $\c T$ is reduced to a smaller complex 
$\tilde{G}$ with tensor $\tilde{\c{T}}$. Repeating this operation many 
times allows one to evaluate tensor traces of arbitrarily large 
complexes. One can therefore compute expectation values of TPS with 
relatively little effort.

In general, the coarse graining transformation is implemented in an 
approximate way. That is, one finds a tensor $\tilde{\c{T}}$ such that 
$\text{wtTr}\left[\c T\otimes \c T \otimes \cdots \right]_G \approx 
\text{wtTr}\left[\tilde{\c{T}}\otimes \tilde{\c{T}} \otimes \cdots 
\right]_{\tilde{G}}$. 
However, as we show below, it can be implemented \emph{exactly} in the case of
the string-net condensed states. Moreover, $\tilde{\c{T}} = \c T$, so that 
the string-net condensed states are fixed points of the TERG 
transformation.

\begin{figure}
\begin{center}
\includegraphics[scale=0.35] {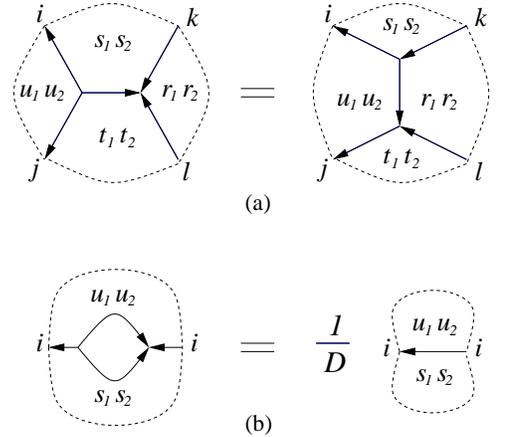}
\end{center}
\caption{The graphic representation of the basic rules for the
double tensor $\c T$. (a) the deformation rule. (b) the reduction
rule.
} \label{Brules}
\end{figure}

\begin{figure}
\begin{center}
\includegraphics[scale=0.35] {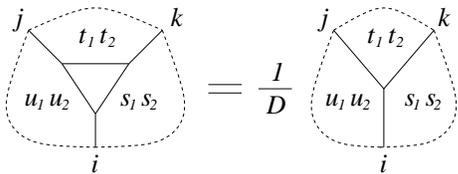}
\end{center}
\caption{ A useful reduction of tensor-complex.
} \label{tBrule}
\end{figure}

To see this, note that the double-$\c T$-tensor has the 
following special properties (see Appendix \ref{rules} for a derivation).
\begin{widetext}
\newtheorem{thm}{Basic rule}
\begin{thm}
The deformation rule of $\c T$:
\begin{eqnarray}
\label{rule1}
\sum_m  \c T_{m,j,i;u_1,s_1,t_1;u_2,s_2,t_2} \c
T_{m^*,k^*,l^*;r_1,t_1,s_1;r_2,t_2,s_2}=\sum_m \c
T_{m,i,k^*;s_1,r_1,u_1;s_2,r_2,u_2} \c
T_{m^*,l^*,j;t_1,u_1,r_1;t_2,u_2,r_2}
\end{eqnarray}
\end{thm}

\begin{thm}
The reduction rule of $\c T$:
\begin{eqnarray}
\label{rule2} \sum_{m,l,t_1,t_2}  a_{t_1} a_{t_2} \c
T_{i,l,m;t_1,s_1,u_1;t_2,s_2,u_2} \c
T_{m^*,l^*,j^*;u_1,s_1,t_1;u_2,s_2,t_2}
=\frac{1}{D}\delta_{ij}\delta_{ {s_2}^* iu_2}\delta_{{s_1}^*iu_1}
\end{eqnarray}
\end{thm}
\end{widetext}
The above two basic rules can be represented graphically
as in Fig. \ref{Brules}. The two basic rules also lead to another
useful property of the double-$\c T$-tensor
as represented by Fig. \ref{tBrule}.

\begin{figure}
\begin{center}
\includegraphics[width=3.5in]
{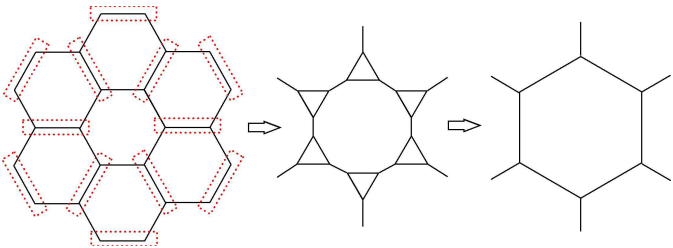}
\end{center}
\caption{The coarse-graining procedure for honeycomb lattice.}
\label{stringnetRG}
\end{figure}

The relations in Fig. \ref{Brules}a and Fig. \ref{tBrule} allow us
to reduce the tensor-complex that represents the norm of the
string-net condensed state to a coarse-grained tensor-complex of
the same shape (see Fig. \ref{stringnetRG}). This coarse-graining 
transformation is exactly the TERG transformation. Since the tensor $\c T$ 
is invariant under such a transformation, we see that the string-net
wave function is a fixed point of the TERG transformation.

\section{Conclusions and discussions}
In conclusion, we found a TPS representation for all the string-net
condensed states.  The local tensors are analogous to local order
parameters in Landau's symmetry broken theory. As a application of
TPS representation, we introduced the TERG transformation and show
that the TPS obtained from the string-net condensed states are fixed
points of the TERG transformation.  In Ref. \onlinecite{GLWtergV}, we
have shown that all the physical measurements, such average energy of
a local Hamiltonian, correlation functions, \etc
can be calculated very efficiently by using the TERG algorithm under
the TPS representations. Thus, TPS representations and TERG
algorithm can capture the long range entanglements in topologically
ordered states. They may be an effective method and a powerful way
to study quantum phases and quantum phase transitions between
different topological orders.

We like to mention that a different wave-function
renormalization procedure is proposed in \Ref{V0705}.  The string-net wave
functions are also fixed point wave functions under such a wave-function
renormalization procedure.

\section{Acknowledgements}
This research is supported by the Foundational Questions Institute
(FQXi) and NSF Grant DMR-0706078.

\appendix

\section{Tensor product state from string-net recoupling rule}
\label{rectps}

To calculate $\Phi^{u,s,t,\cdots}_{i,j,k,\cdots}$, we note that the
coherent states $\ket{u,s,t,\cdots}_\text{coh}$ can be viewed a string-net
state in a fattened lattice (see Fig.
\ref{stringnet}a).\cite{LWstrnet}.  To obtain the string-net states
where strings live on the links, we need to combine the two strings
looping around two adjacent hexagons into a single string on the
link shared by the two hexagons.  We make use of the string-net
recoupling rules,\cite{LWstrnet} see Fig.  \ref{stringnet}, to
represent the coherent states $\ket{u,s,t,\cdots}_\text{coh}$ in terms of
proper orthogonal string states:
\begin{eqnarray}
\ket{u,s,t,\cdots}_\text{coh} &=&\sum_{i,j,k,\cdots} \cdots
F^{u^*u0}_{s^*si}F^{t^*t0}_{u^*uj}F^{s^*s0}_{t^*tk} \cdots \ket{X_b}
\nonumber
\end{eqnarray}
where $X_b$ is the string-net state described in Fig.
\ref{stringnet}b.  Note that on each link we have a factor like
$F^{u^*u0}_{s^*si}$ and we multiply such kind of factors on all
links together. Using $F^{u^*u0}_{s^*si}=\frac{v_i}{v_s v_u}
\delta_{us^*i}$, we can rewrite the above as
\begin{eqnarray}
&&\ket{u,s,t,\cdots}_\text{coh}
\\
&=&\sum_{i,j,k,\cdots} \cdots \frac{v_i}{v_s v_u}\delta_{us^*i}
\frac{v_j}{v_t v_u}\delta_{tu^*j} \frac{v_k}{v_s v_t}\delta_{st^*k}
\cdots \ket{X_b} \nonumber
\end{eqnarray}
Applying the the string-net recoupling rules again, we rewrite the
above as
\begin{eqnarray}
&&\ket{u,s,t,\cdots}_\text{coh}
\\
&=& \sum_{i,j,k,\cdots}\prod_{\text{vertices}}
\frac{\sqrt{v_iv_jv_k}}{v_tv_sv_u}F^{ijk}_{ts^*u}\ket{i,j,k,\cdots}\frac{v_sv_t}{v_k}\nonumber\\
&=& \sum_{i,j,k,\cdots}\prod_{\text{vertices}}
\sqrt{v_iv_jv_k}G^{ijk}_{tsu}\ket{i,j,k,\cdots}
\nonumber\\
\label{fusion}
\end{eqnarray}
where $G^{ijk}_{tsu}=F^{ijk}_{ts^*u}/{v_kv_u}$ is the symmetric $6j$
symbol with full tetrahedral symmetry. Putting Eq. (\ref{fusion})
into Eq. (\ref{strnet}), we finally obtain
\begin{eqnarray}
\label{stringnetwave} \ket{\Psi_{\text{strnet}}} &=&
\sum_{i,j,k,\cdots} \sum_{t,s,u,\cdots} \Big[ \prod_{\text{hexagon}}
a_t  \Big]\times
\\
&&\Big[ \prod_{\text{vertices}} \sqrt{v_iv_jv_k}G^{ijk}_{tsu}
\Big]\ket{i,j,k,\cdots}. \nonumber
\end{eqnarray}
We note that the indices, such as $i,j,k,u,s,t,\cdots$, should be
read from the Fig. \ref{stringnet}. The arrow in(out) on a vertex
will determine that we put $k^*$ or $k$ in the $G$ symbol.  In this
convention, the physical labels in the $G$ symbol are always valued
as $k^*$ on sublattice A(arrow in) and $k$ on sublattice B(arrow
out).

\section{String-net condensed state as a TPS on a tensor-network}
\label{strtnsnet}

\begin{figure}
\begin{center} \includegraphics[width=3.4in] {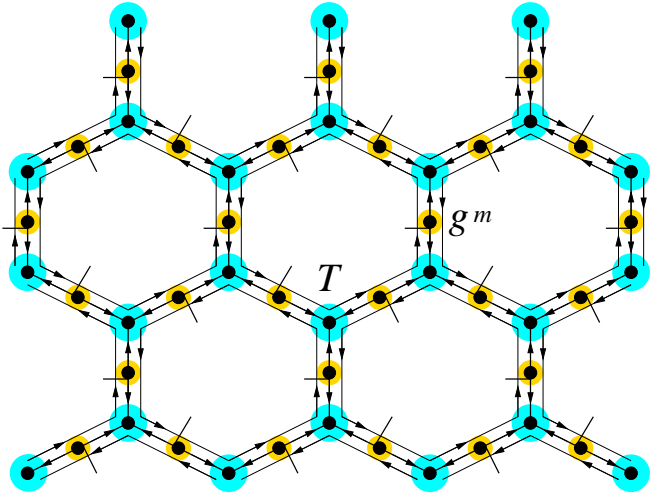}
\end{center}
\caption{
A tensor-network where each tensor has triple-line structure.  The tensor
trace sum over all the indices on the links that connect two dots.
} \label{hextrpln}
\end{figure}

\begin{figure}
\begin{center}
\includegraphics[scale=0.6] {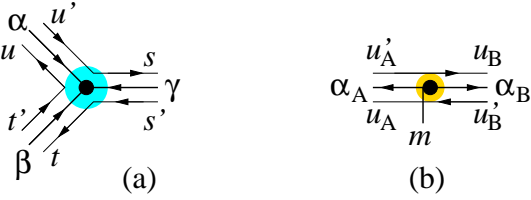}
\end{center}
\caption{
(a) The $T$-tensor,
$T_{u'\al u; t'\bt t; s' \ga s}=
S_{\al\bt\ga; tsu} \del_{ut'}\del_{ts'}\del_{su'}$,
 and (b) the $g^m$-tensor
$g^m_{{u_A' \alpha_A u_A; u_B'\alpha_B u_B}}
= h^m_{\al_A\al_B; u_Au_B} \del_{u_Au_B'}\del_{u_Bu_A'}$,
with a triple-line structure.
} \label{gTtrpln}
\end{figure}

We have expressed the string-net condensed state \eq{stringnetwave1} as a
weighted tensor-trace on the tensor-complex.
In this section, we will show that
the string-net condensed state \eq{stringnetwave1}
can also be
written as a more standard tensor-trace over a tensor-network:
\begin{eqnarray}
\label{strcmpWF}
\ket{\Psi_{\text{strnet}}}=\sum_{m_1,m_2,...}\tx{tTr}[\otimes_v T
\otimes_l g^{m_l}] |m_1,m_2,...\rangle  .
\end{eqnarray}
Here the tensor-network (see Fig. \ref{hextrpln}) is formed by two kinds of
tensors: a $T$-tensor for each vertex and a $g^m$-tensor for each link.  Unlike
the tensors in Fig. \ref{semion}, now
$T$ and $g^m$ have a triple-line structure (see Fig. \ref{gTtrpln}).  The
$T$-tensor on a vertex is given by
\begin{eqnarray}
\label{strnetT1}
T_{u'\alpha u;t'\beta t;s' \gamma s}
&=&  T^0_{\alpha\beta\gamma;tsu}
\delta_{ut^\prime} \delta_{t s^\prime}\delta_{s u^\prime}
\end{eqnarray}
where
\begin{eqnarray}
T^0_{\alpha\beta\gamma;tsu}=(a_t a_sa_u)^{1/6}\sqrt{v_\alpha v_\beta
v_\gamma} G^{\alpha\beta\gamma}_{tsu}.
\end{eqnarray}

As
shown in in Fig. \ref{graphicT0}, $T^0$ can be labeled by three oriented lines
and three faces between them. Notice that $T^0_{\alpha\beta\gamma;tsu}$ has the
cyclic symmetry
\begin{equation*}
T^0_{\alpha\beta\gamma;tsu}=T^0_{\beta\gamma\alpha;sut}
=T^0_{\gamma\alpha\beta;uts}
\end{equation*}
due to the tetrahedron symmetry of $G$ symbol.

The $g^m$-tensor on each link is a projector:
\begin{eqnarray}
g^m_{{u_A' \alpha_A u_A; u_B'\alpha_B u_B}}
&=& h^m_{\al_A\al_B; u_Au_B} \del_{u_Au_B'}\del_{u_Bu_A'},
\nonumber\\
h^m_{\al_A\al_B; u_Au_B} &=&
\delta_{\alpha_A m}\delta_{\al_A\alpha_B^*}
\label{stringnetg1}
\end{eqnarray}
where $m$ is the physical index running from $0$ to $N$ and represents the
$N+1$ different string types (plus the no-string state).  Note that
$u_A,\alpha_A, u_A^\prime$ are indices on the side of the A-sublattice and
$u_B,\alpha_B, u_B^\prime$ are indices on the side of the B-sublattice.
Basically, $g^m$ makes $m=\al_A$ and $\al_A=\al_B^*$.  The corresponding edge
of the hexagon ahs a string of type $m$.  The choice of
$\delta_{\al_A\alpha_B^*}$ in $g^m$ makes the $T$-tensors to be the same on
the A- and B-sublattice.

\begin{figure}[h]
\begin{center}
\includegraphics[width=2in] {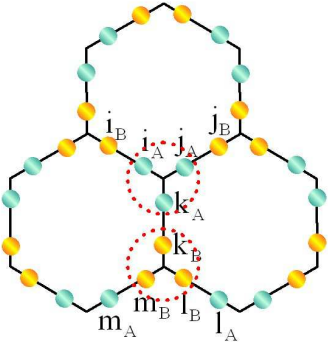}
\end{center}
\caption{The $T$-$g$ form TPS representations for string-net states
can be deformed to standard TPS representations by splitting the
matrix $g^{m}$ into two matrices $g^{m_A}$, $g^{m_B}$ on each
link and recombining three sites around a vertex(the three states
inside a dashed circle) into one site. } 
\label{stringnetsplit}
\end{figure}

Our construction has a slightly different form from the usual TPS
construction, but we can bring our construction into the usual form with a
single set of tensors $T^{\{M\}}$ defined on vertices. This is because the
matrix $g^m$ on each link is basically a projector (with a twist), so that we
can always split it into two matrices $g^{m_A}$ and $g^{m_B}$ (see Fig.
\ref{stringnetsplit}) and associate one with each vertex the link touches.
Doing this for every link in effect displaces the physical degrees of freedom
from the links to the vertices.  This procedure seems to enlarge the Hilbert
space on each link from $H_m$ to $H_{m_A}\otimes H_{m_B}$, however, the
operators $g^{m_A}$ and $g^{m_B}$ impose the constraint $m_A=m_B$ keeping the
physical Hilbert space intact. Grouping the displaced physical degrees of
freedom on each vertex into a new physical variable $M$ we can combine three
sites around each vertex into one site. This is illustrated in Fig.
\ref{stringnetsplit}, where the states in each dashed circle
are labeled by $M$.  With this slight reworking of the
degrees of freedom, the new tensors on sublattice A and B can be expressed as
\begin{eqnarray}
T^{i_Aj_Ak_A}_{A,{u\alpha u^\prime;t\beta t^\prime;s \gamma
s^\prime}} &=&{T^0}_{\alpha\beta\gamma;tsu}^{i_Aj_Ak_A}
\delta_{ut^\prime} \delta_{t s^\prime}\delta_{s
u^\prime}\delta_{\alpha i_A}\delta_{\beta j_A}\delta_{\gamma k_A}
\nonumber\\
T^{k_B l_Bm_B}_{B,u\alpha u^\prime;t\beta t^\prime;s \gamma
s^\prime} &=& {T^0}_{\alpha^*\beta^*\gamma^*;t^\prime s^\prime
u^\prime}^{k_Bl_Bm_B}\delta_{u^\prime t}\delta_{t^\prime
s}\delta_{s^\prime u}\delta_{\alpha
k_B}\delta_{\beta l_B}\delta_{\gamma m_B}, \nonumber\\
\label{stringnetT}
\end{eqnarray}
with
\begin{eqnarray}
{T^0}_{\alpha\beta\gamma;tsu}^{ijk}=(a_ta_sa_u)^{1/6}\sqrt{v_{i}v_{j}v_{k}}G^{\alpha\beta\gamma}_{tsu},
\end{eqnarray}
where $i j k$ are physical indices $M$. Note that the tensors
$T_A^{i_A j_A k_A}$ and $T_B^{k_B l_B m_B}$ still have a triple line
structure. The string-net condensed state now can be rewritten as
\begin{eqnarray}
\ket{\Psi_{\text{strnet}}}=\sum_{M_1,M_2,...}\tx{tTr}\otimes_v
T_{A(B)}^{M_v} |M_1,M_2,...\rangle  . \label{strnetWF2}
\end{eqnarray}

\section{Proof of basic rules of $\c T$}
\label{rules}

The two basic rules can be proved easily by using the Pentagon
Identity\cite{LWstrnet}. First, the Pentagon Identity leads to a
decomposition law for supertensor $\c T$
\begin{eqnarray}
&&\c T_{i,j,k;t_1,s_1,u_1;t_2,s_2,u_2}\\&=&v_iv_jv_k
G^{ijk}_{{t_1}s_1{u_1}}G^{k^*j^*i^*}_{{u_2}{s_2}
{t_2}}\nonumber\\&=&v_iv_jv_k\sum_n d_nG_{{s_1}{u_1}n}^{{u_2}{s_2}^*
i}G_{u_1t_1n}^{{t_2}{u_2}^* j}G_{t_1s_1n}^{{s_2}{t_2}^* k}\nonumber
\end{eqnarray}
Using this expression, it's easy to proof the two basic rules
\eqn{rule1} and \eqn{rule2}. Here we also use the symmetries of $G$
symbol
\begin{eqnarray}
G^{ijk}_{tsu}=G^{kij}_{uts}=G^{ts^*k^*}_{ij^*u^*},
\end{eqnarray}
which can be easily verified from the symmetry of $F$ symbol
\begin{eqnarray}
F^{ijk}_{ts^*u}=F^{kij}_{ut^*s}\frac{v_kv_u}{v_jv_s}=F^{ts^*k^*}_{iju^*}
\end{eqnarray}
and the definition of $G$ symbol.

Then let us proof the deformation rule \eqn{rule1}.
\begin{proof}
\begin{eqnarray}
&&\sum_m  \c T_{m,j,i;u_1,s_1,t_1;u_2,s_2,t_2} \c
T_{m^*,k^*,l^*;r_1,t_1,s_1;r_2,t_2,s_2}\\
&=&v_iv_jv_kv_l\sum_{n,n^\prime,m} d_n d_{n^\prime} d_m
G_{s_1t_1n}^{{t_2}{s_2}^* m}G_{t_1u_1n}^{{u_2}{t_2}^* j}
G_{u_1s_1n}^{{s_2}{u_2}^* i}
\times\nonumber\\&&G_{t_1s_1n^\prime}^{{s_2}{t_2}^*
m^*}G_{s_1r_1n^\prime}^{{r_2}{s_2}^*
k^*}G_{r_1t_1n^\prime}^{{t_2}{r_2}^*l^*}\nonumber\\
&=&v_iv_jv_kv_l\sum_{n,n^\prime} d_n d_{n^\prime} \left(\sum_m d_m
G_{s_1t_1n}^{{t_2}{s_2}^* m}G_{t_1s_1n^\prime}^{{s_2}{t_2}^*
m^*}\right) \times\nonumber\\&&G_{t_1u_1n}^{{u_2}{t_2}^* j}
G_{u_1s_1n}^{{s_2}{u_2}^* i}G_{s_1r_1n^\prime}^{{r_2}{s_2}^*
k^*}G_{r_1t_1n^\prime}^{{t_2}{r_2}^*l^*}\nonumber\\&=&v_iv_jv_kv_l\sum_{n,n^\prime}
d_n d_{n^\prime} \left(\sum_m d_m G_{{t_2}^*{t_1}^*m}^{{s_1}^*{s_2}
n}G_{{t_1}^*{t_2}^*m}^{{s_2}^*{s_1} {n^\prime}^*}\right)
\times\nonumber\\&&G_{t_1u_1n}^{{u_2}{t_2}^* j}
G_{u_1s_1n}^{{s_2}{u_2}^* i}G_{s_1r_1n^\prime}^{{r_2}{s_2}^*
k^*}G_{r_1t_1n^\prime}^{{t_2}{r_2}^*l^*}\nonumber\\&=&
v_iv_jv_kv_l\sum_{n} d_n G_{t_1u_1n}^{{u_2}{t_2}^* j}
G_{u_1s_1n}^{{s_2}{u_2}^* i}G_{s_1r_1n^\prime}^{{r_2}{s_2}^*
k^*}G_{r_1t_1n^\prime}^{{t_2}{r_2}^*l^*}\nonumber
\end{eqnarray}
In the last step we use the simple Pentagon Identity\cite{LWstrnet}
\begin{eqnarray}
\sum_m d_m G_{{t_2}^*{t_1}^*m}^{{s_1}^*{s_2}
n}G_{{t_1}^*{t_2}^*m}^{{s_2}^*{s_1}
{n^\prime}^*}=\frac{\delta_{nn^\prime}}{d_n}\delta_{{s_1}^*s_2n}\delta_{{t_1}^*t_2n^\prime}
\end{eqnarray}
Similarly, we have
\begin{eqnarray}
&&\sum_m \c T_{m,i,k^*;s_1,r_1,u_1;s_2,r_2,u_2} \c
T_{m^*,l^*,j;t_1,u_1,r_1;t_2,u_2,r_2}\\&=& v_iv_jv_kv_l\sum_{n} d_n
 G_{u_1s_1n}^{{s_2}{u_2}^*
i}G_{s_1r_1n^\prime}^{{r_2}{s_2}^*
k^*}G_{r_1t_1n^\prime}^{{t_2}{r_2}^*l^*}G_{t_1u_1n}^{{u_2}{t_2}^* j}
\nonumber
\end{eqnarray}
finally, we have
\begin{eqnarray}
&&\sum_m  \c T_{m,j,i;u_1,s_1,t_1;u_2,s_2,t_2} \c
T_{m^*,k^*,l^*;r_1,t_1,s_1;r_2,t_2,s_2}\\&&=\sum_m \c
T_{m,i,k^*;s_1,r_1,u_1;s_2,r_2,u_2} \c
T_{m^*,l^*,j;t_1,u_1,r_1;t_2,u_2,r_2}\nonumber
\end{eqnarray}
\end{proof}

The reduction rule is also easy to proof by using Pentagon
Identity\cite{LWstrnet}.
\begin{proof}
\begin{eqnarray}
&&\sum_{m,l,t_1,t_2}  a_{t_1} a_{t_2} \c
T_{i,l,m;t_1,s_1,u_1;t_2,s_2,u_2} \c
T_{m^*,l^*,j^*;u_1,s_1,t_1;u_2,s_2,t_2}
\nonumber\\&=&v_iv_j\sum_{n,n^\prime,m,l,t_1,t_2} d_n d_{n^\prime}
d_m d_l a_{t_1} a_{{t_2}^*}\times\nonumber\\
&& G_{s_1u_1n}^{{u_2}{s_2}^* i} G_{u_1t_1n}^{{t_2}{u_2}^*
l}G_{t_1s_1n}^{{s_2}{t_2}^* m}G_{s_1t_1n^\prime}^{{t_2}{s_2}^*
m^*}G_{t_1u_1n^\prime}^{{u_2}{t_2}^*
l^*}G_{u_1s_1n^\prime}^{{s_2}{u_2}^*
j^*}\nonumber\\&=&v_iv_j\sum_{n,n^\prime,t_1,t_2} d_n d_{n^\prime}
a_{t_1} a_{{t_2}^*}G_{s_1u_1n}^{{u_2}{s_2}^*
i}G_{u_1s_1n^\prime}^{{s_2}{u_2}^*
j^*} \times\nonumber\\
&& \left(\sum_l d_l G_{u_1t_1n}^{{t_2}{u_2}^*
l}G_{t_1u_1n^\prime}^{{u_2}{t_2}^* l^*}\right)\left(\sum_m
d_mG_{t_1s_1n}^{{s_2}{t_2}^* m}G_{s_1t_1n^\prime}^{{t_2}{s_2}^*
m^*}\right)\nonumber\\&=& v_iv_j\sum_{n,t_1,t_2} \delta_{{t_1}^* t_2
n}a_{t_1} a_{t_2} G_{s_1u_1n}^{{u_2}{s_2}^* i}
G_{u_1s_1n}^{{s_2}{u_2}^* j^*}\nonumber\\&=& v_iv_j\sum_{n} a_n
G_{s_1u_1n}^{{u_2}{s_2}^* i} G_{u_1s_1n}^{{s_2}{u_2}^* j^*}
\nonumber\\&=& \frac{v_iv_j}{D}\sum_{n} d_n
G_{s_1u_1n}^{{u_2}{s_2}^* i} G_{u_1s_1n}^{{s_2}{u_2}^* j^*}
\nonumber\\&=&\frac{1}{D}\delta_{ij}\delta_{ {s_2}^*
iu_2}\delta_{{s_1}^*iu_1}
\end{eqnarray}
\end{proof}
Notice in the third line, we use the fact that
$a_{t_2}=a_{{t_2}^*}$.


\begin{thebibliography}{35}
\expandafter\ifx\csname natexlab\endcsname\relax\def\natexlab#1{#1}\fi
\expandafter\ifx\csname bibnamefont\endcsname\relax
  \def\bibnamefont#1{#1}\fi
\expandafter\ifx\csname bibfnamefont\endcsname\relax
  \def\bibfnamefont#1{#1}\fi
\expandafter\ifx\csname citenamefont\endcsname\relax
  \def\citenamefont#1{#1}\fi
\expandafter\ifx\csname url\endcsname\relax
  \def\url#1{\texttt{#1}}\fi
\expandafter\ifx\csname urlprefix\endcsname\relax\def\urlprefix{URL }\fi
\providecommand{\bibinfo}[2]{#2}
\providecommand{\eprint}[2][]{\url{#2}}

\bibitem[{\citenamefont{Landau}(1937)}]{L3726}
\bibinfo{author}{\bibfnamefont{L.~D.} \bibnamefont{Landau}},
  \bibinfo{journal}{Phys. Z. Sowjetunion} \textbf{\bibinfo{volume}{11}},
  \bibinfo{pages}{26} (\bibinfo{year}{1937}).

\bibitem[{\citenamefont{Tsui et~al.}(1982)\citenamefont{Tsui, Stormer, and
  Gossard}}]{TSG8259}
\bibinfo{author}{\bibfnamefont{D.~C.} \bibnamefont{Tsui}},
  \bibinfo{author}{\bibfnamefont{H.~L.} \bibnamefont{Stormer}},
  \bibnamefont{and} \bibinfo{author}{\bibfnamefont{A.~C.}
  \bibnamefont{Gossard}}, \bibinfo{journal}{Phys. Rev. Lett.}
  \textbf{\bibinfo{volume}{48}}, \bibinfo{pages}{1559} (\bibinfo{year}{1982}).

\bibitem[{\citenamefont{Wen}(1990)}]{Wrig}
\bibinfo{author}{\bibfnamefont{X.-G.} \bibnamefont{Wen}},
  \bibinfo{journal}{Int. J. Mod. Phys. B} \textbf{\bibinfo{volume}{4}},
  \bibinfo{pages}{239} (\bibinfo{year}{1990}).

\bibitem[{\citenamefont{Wen and Wu}(1993)}]{WWtran}
\bibinfo{author}{\bibfnamefont{X.-G.} \bibnamefont{Wen}} \bibnamefont{and}
  \bibinfo{author}{\bibfnamefont{Y.-S.} \bibnamefont{Wu}},
  \bibinfo{journal}{Phys. Rev. Lett.} \textbf{\bibinfo{volume}{70}},
  \bibinfo{pages}{1501} (\bibinfo{year}{1993}).

\bibitem[{\citenamefont{Senthil et~al.}(1999)\citenamefont{Senthil, Marston,
  and Fisher}}]{SMF9945}
\bibinfo{author}{\bibfnamefont{T.}~\bibnamefont{Senthil}},
  \bibinfo{author}{\bibfnamefont{J.~B.} \bibnamefont{Marston}},
  \bibnamefont{and} \bibinfo{author}{\bibfnamefont{M.~P.~A.}
  \bibnamefont{Fisher}}, \bibinfo{journal}{Phys. Rev. B}
  \textbf{\bibinfo{volume}{60}}, \bibinfo{pages}{4245} (\bibinfo{year}{1999}).

\bibitem[{\citenamefont{Wen}(2000)}]{Wctpt}
\bibinfo{author}{\bibfnamefont{X.-G.} \bibnamefont{Wen}},
  \bibinfo{journal}{Phys. Rev. Lett.} \textbf{\bibinfo{volume}{84}},
  \bibinfo{pages}{3950} (\bibinfo{year}{2000}).

\bibitem[{\citenamefont{Wen}(2002{\natexlab{a}})}]{Wqoslpub}
\bibinfo{author}{\bibfnamefont{X.-G.} \bibnamefont{Wen}},
  \bibinfo{journal}{Phys. Rev. B} \textbf{\bibinfo{volume}{65}},
  \bibinfo{pages}{165113} (\bibinfo{year}{2002}{\natexlab{a}}).

\bibitem[{\citenamefont{Senthil et~al.}(2004)\citenamefont{Senthil, Balents,
  Sachdev, Vishwanath, and Fisher}}]{SBS0407}
\bibinfo{author}{\bibfnamefont{T.}~\bibnamefont{Senthil}},
  \bibinfo{author}{\bibfnamefont{L.}~\bibnamefont{Balents}},
  \bibinfo{author}{\bibfnamefont{S.}~\bibnamefont{Sachdev}},
  \bibinfo{author}{\bibfnamefont{A.}~\bibnamefont{Vishwanath}},
  \bibnamefont{and} \bibinfo{author}{\bibfnamefont{M.~P.~A.}
  \bibnamefont{Fisher}}, \bibinfo{journal}{Physical Review B}
  \textbf{\bibinfo{volume}{70}}, \bibinfo{pages}{144407}
  (\bibinfo{year}{2004}).

\bibitem[{\citenamefont{Senthil}(2004)}]{SenthilQCP}
\bibinfo{author}{\bibfnamefont{T.}~\bibnamefont{Senthil}},
  \bibinfo{journal}{Science} \textbf{\bibinfo{volume}{303}},
  \bibinfo{pages}{1490} (\bibinfo{year}{2004}).

\bibitem[{\citenamefont{Read and Sachdev}(1991)}]{RS9173}
\bibinfo{author}{\bibfnamefont{N.}~\bibnamefont{Read}} \bibnamefont{and}
  \bibinfo{author}{\bibfnamefont{S.}~\bibnamefont{Sachdev}},
  \bibinfo{journal}{Phys. Rev. Lett.} \textbf{\bibinfo{volume}{66}},
  \bibinfo{pages}{1773} (\bibinfo{year}{1991}).

\bibitem[{\citenamefont{Wen}(1991)}]{Wsrvb}
\bibinfo{author}{\bibfnamefont{X.-G.} \bibnamefont{Wen}},
  \bibinfo{journal}{Phys. Rev. B} \textbf{\bibinfo{volume}{44}},
  \bibinfo{pages}{2664} (\bibinfo{year}{1991}).

\bibitem[{\citenamefont{Senthil and Fisher}(2000)}]{SF0050}
\bibinfo{author}{\bibfnamefont{T.}~\bibnamefont{Senthil}} \bibnamefont{and}
  \bibinfo{author}{\bibfnamefont{M.~P.~A.} \bibnamefont{Fisher}},
  \bibinfo{journal}{Phys. Rev. B} \textbf{\bibinfo{volume}{62}},
  \bibinfo{pages}{7850} (\bibinfo{year}{2000}).

\bibitem[{\citenamefont{Moessner and Sondhi}(2001)}]{MS0181}
\bibinfo{author}{\bibfnamefont{R.}~\bibnamefont{Moessner}} \bibnamefont{and}
  \bibinfo{author}{\bibfnamefont{S.~L.} \bibnamefont{Sondhi}},
  \bibinfo{journal}{Phys. Rev. Lett.} \textbf{\bibinfo{volume}{86}},
  \bibinfo{pages}{1881} (\bibinfo{year}{2001}).

\bibitem[{\citenamefont{Sachdev and Park}(2002)}]{SP0258}
\bibinfo{author}{\bibfnamefont{S.}~\bibnamefont{Sachdev}} \bibnamefont{and}
  \bibinfo{author}{\bibfnamefont{K.}~\bibnamefont{Park}},
  \bibinfo{journal}{Annals of Physics (N.Y.)} \textbf{\bibinfo{volume}{298}},
  \bibinfo{pages}{58} (\bibinfo{year}{2002}).

\bibitem[{\citenamefont{Misguich et~al.}(2002)\citenamefont{Misguich, Serban,
  and Pasquier}}]{MSP0202}
\bibinfo{author}{\bibfnamefont{G.}~\bibnamefont{Misguich}},
  \bibinfo{author}{\bibfnamefont{D.}~\bibnamefont{Serban}}, \bibnamefont{and}
  \bibinfo{author}{\bibfnamefont{V.}~\bibnamefont{Pasquier}},
  \bibinfo{journal}{Phys. Rev. Lett.} \textbf{\bibinfo{volume}{89}},
  \bibinfo{pages}{137202} (\bibinfo{year}{2002}).

\bibitem[{\citenamefont{Balents et~al.}(2002)\citenamefont{Balents, Fisher, and
  Girvin}}]{BFG0212}
\bibinfo{author}{\bibfnamefont{L.}~\bibnamefont{Balents}},
  \bibinfo{author}{\bibfnamefont{M.~P.~A.} \bibnamefont{Fisher}},
  \bibnamefont{and} \bibinfo{author}{\bibfnamefont{S.~M.}
  \bibnamefont{Girvin}}, \bibinfo{journal}{Phys. Rev. B}
  \textbf{\bibinfo{volume}{65}}, \bibinfo{pages}{224412}
  (\bibinfo{year}{2002}).

\bibitem[{\citenamefont{Wen}(2003)}]{Wqoexct}
\bibinfo{author}{\bibfnamefont{X.-G.} \bibnamefont{Wen}},
  \bibinfo{journal}{Phys. Rev. Lett.} \textbf{\bibinfo{volume}{90}},
  \bibinfo{pages}{016803} (\bibinfo{year}{2003}).

\bibitem[{\citenamefont{Kitaev}(2003)}]{K032}
\bibinfo{author}{\bibfnamefont{A.~Y.} \bibnamefont{Kitaev}},
  \bibinfo{journal}{Ann. Phys. (N.Y.)} \textbf{\bibinfo{volume}{303}},
  \bibinfo{pages}{2} (\bibinfo{year}{2003}).

\bibitem[{\citenamefont{Ioffe et~al.}(2002)\citenamefont{Ioffe, Feigel'man,
  Ioselevich, Ivanov, Troyer, and Blatter}}]{IFI0203}
\bibinfo{author}{\bibfnamefont{L.~B.} \bibnamefont{Ioffe}},
  \bibinfo{author}{\bibfnamefont{M.~V.} \bibnamefont{Feigel'man}},
  \bibinfo{author}{\bibfnamefont{A.}~\bibnamefont{Ioselevich}},
  \bibinfo{author}{\bibfnamefont{D.}~\bibnamefont{Ivanov}},
  \bibinfo{author}{\bibfnamefont{M.}~\bibnamefont{Troyer}}, \bibnamefont{and}
  \bibinfo{author}{\bibfnamefont{G.}~\bibnamefont{Blatter}},
  \bibinfo{journal}{Nature} \textbf{\bibinfo{volume}{415}},
  \bibinfo{pages}{503} (\bibinfo{year}{2002}).

\bibitem[{\citenamefont{Freedman et~al.}(2004)\citenamefont{Freedman, Nayak,
  Shtengel, Walker, and Wang}}]{FNS0428}
\bibinfo{author}{\bibfnamefont{M.}~\bibnamefont{Freedman}},
  \bibinfo{author}{\bibfnamefont{C.}~\bibnamefont{Nayak}},
  \bibinfo{author}{\bibfnamefont{K.}~\bibnamefont{Shtengel}},
  \bibinfo{author}{\bibfnamefont{K.}~\bibnamefont{Walker}}, \bibnamefont{and}
  \bibinfo{author}{\bibfnamefont{Z.}~\bibnamefont{Wang}},
  \bibinfo{journal}{Ann. Phys. (NY)} \textbf{\bibinfo{volume}{310}},
  \bibinfo{pages}{428} (\bibinfo{year}{2004}).

\bibitem[{\citenamefont{Levin and Wen}(2005{\natexlab{a}})}]{LWstrnet}
\bibinfo{author}{\bibfnamefont{M.}~\bibnamefont{Levin}} \bibnamefont{and}
  \bibinfo{author}{\bibfnamefont{X.-G.} \bibnamefont{Wen}},
  \bibinfo{journal}{Phys. Rev. B} \textbf{\bibinfo{volume}{71}},
  \bibinfo{pages}{045110} (\bibinfo{year}{2005}{\natexlab{a}}).

\bibitem[{\citenamefont{Levin and Wen}(2005{\natexlab{b}})}]{LWuni}
\bibinfo{author}{\bibfnamefont{M.~A.} \bibnamefont{Levin}} \bibnamefont{and}
  \bibinfo{author}{\bibfnamefont{X.-G.} \bibnamefont{Wen}},
  \bibinfo{journal}{Rev. Mod. Phys.} \textbf{\bibinfo{volume}{77}},
  \bibinfo{pages}{871} (\bibinfo{year}{2005}{\natexlab{b}}).

\bibitem[{\citenamefont{Kitaev and Preskill}(2006)}]{KP0604}
\bibinfo{author}{\bibfnamefont{A.}~\bibnamefont{Kitaev}} \bibnamefont{and}
  \bibinfo{author}{\bibfnamefont{J.}~\bibnamefont{Preskill}},
  \bibinfo{journal}{Phys. Rev. Lett.} \textbf{\bibinfo{volume}{96}},
  \bibinfo{pages}{110404} (\bibinfo{year}{2006}).

\bibitem[{\citenamefont{Levin and Wen}(2006)}]{LWtopent}
\bibinfo{author}{\bibfnamefont{M.}~\bibnamefont{Levin}} \bibnamefont{and}
  \bibinfo{author}{\bibfnamefont{X.-G.} \bibnamefont{Wen}},
  \bibinfo{journal}{Phys. Rev. Lett.} \textbf{\bibinfo{volume}{96}},
  \bibinfo{pages}{110405} (\bibinfo{year}{2006}).

\bibitem[{\citenamefont{Verstraete and Cirac}()}]{FrankPEPS1}
\bibinfo{author}{\bibfnamefont{F.}~\bibnamefont{Verstraete}} \bibnamefont{and}
  \bibinfo{author}{\bibfnamefont{J.~I.} \bibnamefont{Cirac}},
  \bibinfo{howpublished}{cond-mat/0407066}.

\bibitem[{\citenamefont{Vidal}(2007)}]{V0705}
\bibinfo{author}{\bibfnamefont{G.}~\bibnamefont{Vidal}},
  \bibinfo{journal}{Phys. Rev. Lett.} \textbf{\bibinfo{volume}{99}},
  \bibinfo{pages}{220405} (\bibinfo{year}{2007}).

\bibitem[{\citenamefont{Aguado and Vidal}(2008{\natexlab{a}})}]{AV0804}
\bibinfo{author}{\bibfnamefont{M.}~\bibnamefont{Aguado}} \bibnamefont{and}
  \bibinfo{author}{\bibfnamefont{G.}~\bibnamefont{Vidal}},
  \bibinfo{journal}{Phys. Rev. Lett.} \textbf{\bibinfo{volume}{100}},
  \bibinfo{pages}{070404} (\bibinfo{year}{2008}{\natexlab{a}}).

\bibitem[{\citenamefont{Klumper et~al.}(1991)\citenamefont{Klumper,
  Schadschneider, and Zittartz}}]{KSZ9155}
\bibinfo{author}{\bibfnamefont{A.}~\bibnamefont{Klumper}},
  \bibinfo{author}{\bibfnamefont{A.}~\bibnamefont{Schadschneider}},
  \bibnamefont{and} \bibinfo{author}{\bibfnamefont{J.}~\bibnamefont{Zittartz}},
  \bibinfo{journal}{J. Phys. A} \textbf{\bibinfo{volume}{24}},
  \bibinfo{pages}{L955} (\bibinfo{year}{1991}).

\bibitem[{\citenamefont{Perez-Garcia et~al.}(2007)\citenamefont{Perez-Garcia,
  Verstraete, Wolf, and Cirac}}]{PVW0701}
\bibinfo{author}{\bibfnamefont{D.}~\bibnamefont{Perez-Garcia}},
  \bibinfo{author}{\bibfnamefont{F.}~\bibnamefont{Verstraete}},
  \bibinfo{author}{\bibfnamefont{M.}~\bibnamefont{Wolf}}, \bibnamefont{and}
  \bibinfo{author}{\bibfnamefont{J.}~\bibnamefont{Cirac}},
  \bibinfo{journal}{Quantum Inf. Comput.} \textbf{\bibinfo{volume}{7}},
  \bibinfo{pages}{401} (\bibinfo{year}{2007}).

\bibitem[{\citenamefont{White}(1992)}]{W9263}
\bibinfo{author}{\bibfnamefont{S.~R.} \bibnamefont{White}},
  \bibinfo{journal}{Phys. Rev. Lett.} \textbf{\bibinfo{volume}{69}},
  \bibinfo{pages}{2863} (\bibinfo{year}{1992}).

\bibitem[{\citenamefont{Wen}(2002{\natexlab{b}})}]{W0275}
\bibinfo{author}{\bibfnamefont{X.-G.} \bibnamefont{Wen}},
  \bibinfo{journal}{Physics Letters A} \textbf{\bibinfo{volume}{300}},
  \bibinfo{pages}{175} (\bibinfo{year}{2002}{\natexlab{b}}).

\bibitem[{\citenamefont{Verstraete et~al.}(2006)\citenamefont{Verstraete, Wolf,
  Perez-Garcia, and Cirac}}]{FrankPEPS2}
\bibinfo{author}{\bibfnamefont{F.}~\bibnamefont{Verstraete}},
  \bibinfo{author}{\bibfnamefont{M.~M.} \bibnamefont{Wolf}},
  \bibinfo{author}{\bibfnamefont{D.}~\bibnamefont{Perez-Garcia}},
  \bibnamefont{and} \bibinfo{author}{\bibfnamefont{J.~I.} \bibnamefont{Cirac}},
  \bibinfo{journal}{Phys. Rev. Lett.} \textbf{\bibinfo{volume}{96}},
  \bibinfo{pages}{220601} (\bibinfo{year}{2006}).

\bibitem[{\citenamefont{Aguado and Vidal}(2008{\natexlab{b}})}]{VidalMERA}
\bibinfo{author}{\bibfnamefont{M.}~\bibnamefont{Aguado}} \bibnamefont{and}
  \bibinfo{author}{\bibfnamefont{G.}~\bibnamefont{Vidal}},
  \bibinfo{journal}{Phys. Rev. Lett.} \textbf{\bibinfo{volume}{100}},
  \bibinfo{pages}{070404} (\bibinfo{year}{2008}{\natexlab{b}}).

\bibitem[{\citenamefont{Levin and Nave}(2007)}]{LevinTRG}
\bibinfo{author}{\bibfnamefont{M.}~\bibnamefont{Levin}} \bibnamefont{and}
  \bibinfo{author}{\bibfnamefont{C.~P.} \bibnamefont{Nave}},
  \bibinfo{journal}{Phys. Rev. Lett.} \textbf{\bibinfo{volume}{99}},
  \bibinfo{pages}{120601} (\bibinfo{year}{2007}).

\bibitem[{\citenamefont{Gu et~al.}(2008)\citenamefont{Gu, Levin, and
  Wen}}]{GLWtergV}
\bibinfo{author}{\bibfnamefont{Z.-C.} \bibnamefont{Gu}},
  \bibinfo{author}{\bibfnamefont{M.}~\bibnamefont{Levin}}, \bibnamefont{and}
  \bibinfo{author}{\bibfnamefont{X.-G.} \bibnamefont{Wen}},
  \bibinfo{journal}{arXiv:0806.3509}  (\bibinfo{year}{2008}).

\end{thebibliography}

\end{document}